\documentclass[twocolumn,aps,prl,preprintnumbers]{revtex4}
\usepackage{bbm}
\usepackage[latin9]{inputenc}
\usepackage{amsmath}
\usepackage{amssymb}
\usepackage{graphicx}
\usepackage{mathrsfs}
\usepackage{amsfonts}
\usepackage{amsthm}
\usepackage{color}
\usepackage{txfonts}
\usepackage[colorlinks=true,citecolor=blue,linkcolor=blue,urlcolor=blue,anchorcolor=blue]{hyperref}%
\usepackage{pifont}
\hypersetup{colorlinks=true,citecolor=blue,linkcolor=blue,urlcolor=blue}
\providecommand{\U}[1]{\protect\rule{.1in}{.1in}}
\makeatletter
\@ifundefined{textcolor}{}
{
	\definecolor{BLACK}{gray}{0}
	\definecolor{WHITE}{gray}{1}
	\definecolor{RED}{rgb}{1,0,0}
	\definecolor{GREEN}{rgb}{0,1,0}
	\definecolor{BLUE}{rgb}{0,0,1}
	\definecolor{CYAN}{cmyk}{1,0,0,0}
	\definecolor{MAGENTA}{cmyk}{0,1,0,0}
	\definecolor{YELLOW}{cmyk}{0,0,1,0}
}

\makeatother
\begin{document}

\title{Magnon spin transport through atomic ferrimagnetic domain walls}

\author{Zhaozhuo Zeng}
\author{Peng Yan}
\email{yan@uestc.edu.cn}
\affiliation{School of Physics and State Key Laboratory of Electronic Thin Films and Integrated Devices, University of Electronic Science and Technology of China, Chengdu 611731, China}

\begin{abstract}
It is a well-established notion that the spin of a magnon should be flipped when it passes through a $180^{\circ}$ domain wall (DW) in both ferromagnets and antiferromagnets, while the magnon spin transport through ferrimagnetic DW is still elusive. In this work, we report that the magnon preserves its spin after the transmission through an atomically sharp DW in ferrimagnets, due to the intriguing interband magnon scattering at the domain interface. This finding may provide significant insight to resolve the puzzling insensitivity of magnon spin diffusion to the $180^{\circ}$ ferrimagnetic DWs observed by recent experiments. Our results reveal the unique role of ferrimagnetic DWs in manipulating the magnon spin and may facilitate the design of novel magnonic devices based on ferrimagnets.
\end{abstract}

\maketitle

Magnons (or spin waves) are quasiparticles associated with wave-like disturbances in ordered magnets, and are able to carry both linear and angular momentum. The rise of the emerging magnonics is largely due to the low energy dissipation and long coherence length of magnons \cite{serga2010yig, chumak2015magnon, yu2021magnetic, barman20212021,yuan2022quantum}. One fundamental issue is to understand the interaction between magnons and magnetic textures \cite{yan2011all, wang2012domain, yu2016magnetic, hamalainen2018control, oh2019bidirectional, han2019mutual, liang2022nonreciprocal, lan2017antiferromagnetic, faridi2022atomic,iwasaki2014theory,schutte2014magnon,kim2019tunable,wang2021magnonic,li2022interaction}. Magnetic domain wall (DW), the transition region separating two magnetic domains, is a prominent example of spin textures that are promising for information industry \cite{schryer1974motion,wang2018theory}. It has been known that a magnon will transfer its angular momentum when it passes through the DW, leading to the so-called magnonic spin transfer torque \cite{yan2011all}. When the DW is wide, a continuum model is convenient to describe the magnon transport since it allows analytical solutions. For a narrow DW, however, the continuum model does not suffice to characterize the spin texture and an atomistic approach is demanded to interpret, for instance, the atomic DW pinning \cite{novoselov2003subatomic}, the strong magnon reflection \cite{yan2012magnonic}, and the magnonic Cherenkov emission \cite{Yang2019}. Narrow DWs have been observed in antiferromagnets as well, like CuMnAs \cite{krizek2022atomically} and $\mathrm{FePS_3}$ \cite{lee2023giant}. As their ferromagnetic counterpart, the spin of a magnon undergoes a sign flipping as it traverses antiferromagnetic DWs \cite{tveten2014antiferromagnetic,kim2014propulsion}.

Ferrimagnets (FiMs), which exhibit antiferromagnetic coupling but non-zero net magnetization, can be readily manipulated for ultrafast devices \cite{kim2022ferrimagnetic, zhang2023ferrimagnets}. Due to different angular momentum densities of two sublattices in FiMs, there are two non-degenerate circularly polarized states with contrast magnon dispersions even without the external magnetic field, providing a new degree of spin freedom \cite{oh2017coherent, kim2021current, kim2020distinct,okamoto2020flipping}. Very recently, a puzzling insensitivity of magnon spin diffusion in multidomain ferrimagnet has been reported \cite{li2023puzzling}. Following the conventional wisdom for both ferromagnets and antiferromagnets that magnons flip their spin after passing through the DW, one may expect that the angular momentum carried by magnons in multi-domain state should quickly decay to zero because of the cancellation effect in opposite domains. However, non-local measurements show almost the same signal strength as that in a single domain state \cite{li2023puzzling}. This thus bring about a critical question: Do magnon reverse or reserve its spin after passing through the DW in FiMs?

In this work, we theoretically investigate the angular momentum transport of magnons through ferrimagnetic DWs in atomic scales. For wide DWs, magnons indeed switch their spin and meanwhile reserve the linear momentum after passing through the DW, which echoes the case for both ferromagnet and antiferromagnet. Interestingly, we find that, for atomically sharp DWs, the magnon spin remains after the transmission due to the intriguing interband magnon scattering, which is accompanied by a linear momentum jump. Our results advance the understanding of the interaction between magnons and magnetic textures, and may offer stimulating insight to clarify the aforementioned puzzling insensitivity of magnon spin diffusion to ferrimagnetic DWs.

\begin{figure}[t]
	\includegraphics[width=1\columnwidth]{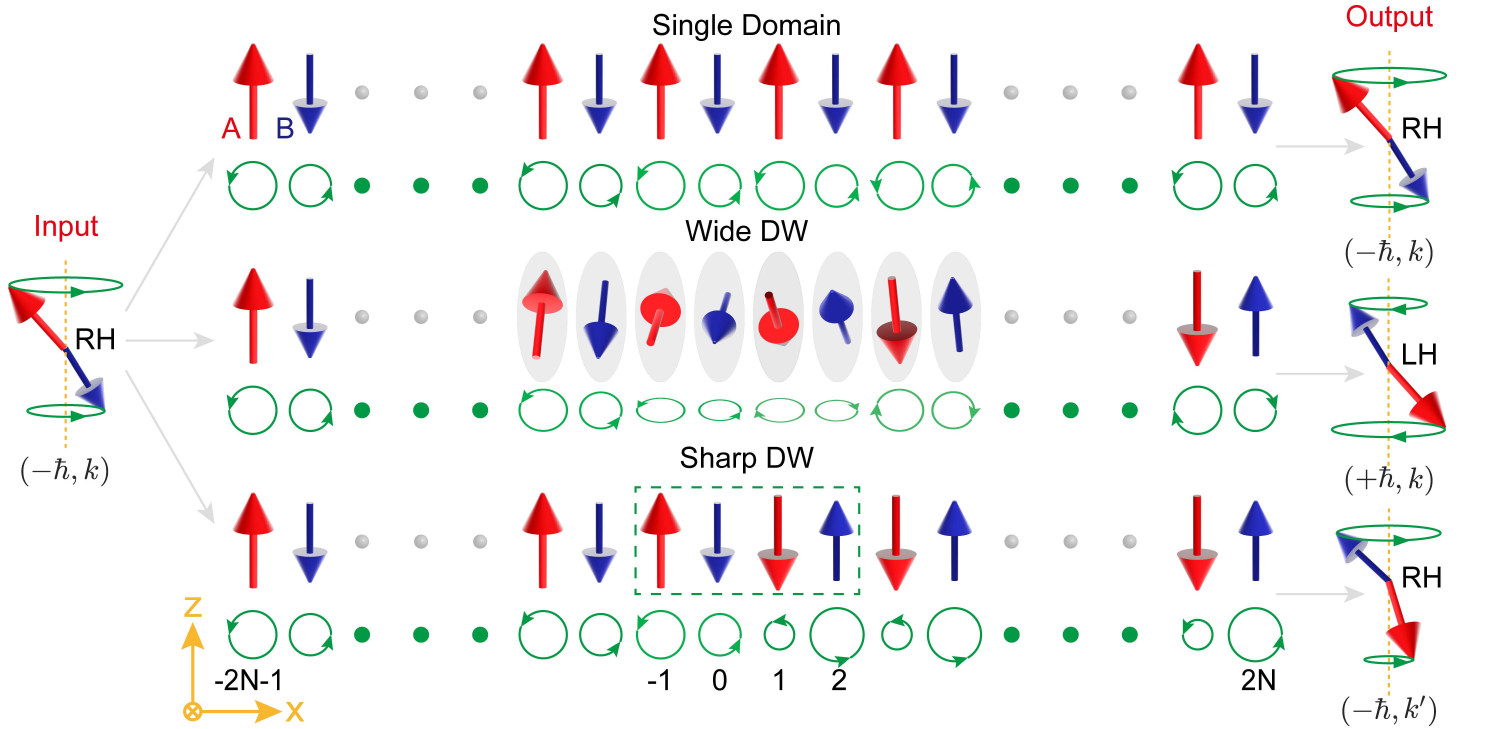}%
	\caption{\label{Fig:1}Spin wave transport through a single domain, wide DW, and sharp DW in ferrimagnets, respectively. Green circles are the top view of the magnetization precession. $A$ and $B$ label two sublattices. $\pm\hbar$ represents the magnon spin. $k$ and $k^{\prime}$ indicate the magnon wave vector.}
\end{figure}

Let's consider the following one-dimensional atomistic spin Hamiltonian
\begin{equation}
    \mathcal{H} = J {\sum\limits_{n}} {\bf{S}}_{n} \cdot {\bf{S}}_{n+1} - K {\sum\limits_{n}} ({S}_{n}^{z})^{2}, \label{eq:1}
\end{equation}
where $ J > 0 $ and $ K >0 $ represent the antiferromagnetic exchange couple and easy-axis magnetic anisotropy, respectively. Due to the competition between these two energies and the boundary condition, there are three kinds of magnetization profiles, i.e., a single domain, a wide DW, and an atomically sharp DW, as shown in Fig. \ref{Fig:1}. It is noted that the width of DW is determined by the radio $ K/J $ (see below).

The equation of motion for each spin with index $ n $ can be expressed as $s_{l} \partial_t {\bf{S}}_{n} = -{\bf{S}}_{n} \times {\bf{H}}_{\mathrm{eff},n}$, where $ s_{l} = \mu_{l} / \gamma_{l}  $ $ (l = \mathrm{A,B}) $ is the angular momentum density, $ \mu_{l} $ is the magnetic moment, $ \gamma_{l} $ is the gyromagnetic ratio, and  $ {\bf{H}}_{\mathrm{eff},n} = -\delta \mathcal{H}_{n} / \delta {\bf{S}}_{n} $ is the effective field. When the index $ n $ is odd (even), $ l $ represents A (B) sublattice. It is worth mentioning that $ s_{\mathrm{A}} \neq s_{\mathrm{B}} $ in general for FiMs, because of which the degeneracy of $ \pm\hbar $ magnons is broken even without the external magnetic field.

We first linearize the equation of motion to obtain the dispersion relation of magnons. To this end, we express the spin $ ( S_{n}^{x}, S_{n}^{y}, S_{n}^{z} ) $ in local frames as
\begin{equation}
\left(
\begin{array}{c}
S_{n}^{x}\\
S_{n}^{y}\\
S_{n}^{z}
\end{array}
\right )
=
\left(
\begin{array}{cccc}
1 & 0 & 0\\
0 & \cos\theta_{n} & \sin\theta_{n}\\
0 & -\sin\theta_{n} & \cos\theta_{n}
\end{array}
\right )
\left(
\begin{array}{c}
S_{n}^{X}\\
S_{n}^{Y}\\
S_{n}^{Z}
\end{array}
\right ), \label{eq:2}
\end{equation}
where $ S_{n}^{Z} \approx 1, |S_{n}^{X}|, |S_{n}^{Y}| \ll 1$, and $\theta_n$ is the equilibrium deflection angle of the magnetization ${\bf{S}}_n $ from the $z$-axis, satisfying
\begin{equation}
	J\sin(\theta_{n-1}-\theta_{n}) + J\sin(\theta_{n+1}-\theta_{n}) + K\sin(2\theta_{n}) = 0. \label{eq:3}
\end{equation}
We then obtain
\begin{equation}
	\begin{aligned}
	s_{l} \partial_t S_{n}^{X} =& \left[Jc_{n-1} + Jc_{n} - 2K(\cos^{2}\theta_{n}-\sin^{2}\theta_{n})\right] S_{n}^{Y}\\
	 &- Jc_{n-1} S_{n-1}^{Y} - Jc_{n} S_{n+1}^{Y},\\
	s_{l} \partial_t S_{n}^{Y} =&(-Jc_{n-1} - Jc_{n} - 2K\cos^{2}\theta_{n}) S_{n}^{X}\\
	 &+ JS_{n-1}^{X} + JS_{n+1}^{X},
	\end{aligned} \label{eq:4}
\end{equation}
where $ c_{n} = \cos(\theta_{n}-\theta_{n+1}) $. 
To facilitate the analysis,  we define wave functions of right-handed (RH) and left-handed (LH) precession of local magnetic moments as $ \psi_{n}^{\pm} = S_{n}^{X} \pm iS_{n}^{Y} = \Xi_{s_{l}}^{\pm}\text{exp}[i (\omega t-kna)]$, where $ \Xi_{s_{l}}^{+} $ and $ \Xi_{s_{l}}^{-} $ are the amplitudes of RH and LH precessions, respectively, $ k $ is the wave vector, $a$ is the lattice constant, and $\omega/2\pi$ is the frequency.

For a single domain shown in the top panel of Fig.~\ref{Fig:1}, where $ \theta_{2n-1} (\theta_{2n}) = 0 (\pi) $ with the integer $ n $ from $ -N $ to $ N $, we derive the dispersion and amplitude ratio for the RH (LH) magnon \cite{SM}
\begin{equation}
\omega_{\pm} = \frac{ \mp \zeta + \sqrt{\zeta^{2} + \eta}}{s_\mathrm{A}s_\mathrm{B}}, \label{eq:5}
\end{equation}
\begin{equation}
\rho^\mathrm{\pm} = \frac{\Xi_{s_\mathrm{A}}^{\pm}}{\Xi_{s_\mathrm{B}}^{\mp}} = - \frac{2J\cos ka}{2J + 2K \mp s_\mathrm{A} \omega_\mathrm{\pm}}, \label{eq:6}
\end{equation}
where the RH (LH) magnon takes the negative (positive) sign, $ \zeta = (s_\mathrm{A}-s_\mathrm{B})(J+K) $ and $ \eta = 4s_\mathrm{A}s_\mathrm{B} [(J+K)^{2} - J^{2} \cos^{2}(ka)] $. 

\textit{The wide DW.---}When $K \ll J$, we have a wide FiM DW. The atomistic model then can be reduced to a continuum form. We then obtain the equation of motion of the N\'{e}el vector $ {\bf{n}} = \left({\bf{S}}_\mathrm{A}-{\bf{S}}_\mathrm{B}\right)/2 $ \cite{kim2022ferrimagnetic}
\begin{equation}
	\partial_t (\sigma_s {\bf{n}} \times \dot{{\bf{n}}} - \delta_s {\bf{n}}) = \nabla\cdot {\bf{J}} +K_u n_z {\bf{n}} \times \hat{z}, \label{eq:7}
\end{equation}
where $ {\bf{J}} = A {\bf{n}} \times \nabla {\bf{n}} $ is the magnon spin current \cite{tveten2014antiferromagnetic,kim2014propulsion}, $ \sigma_s = s^2 a^3 / 4J $, $ \delta_s = s_\mathrm{A} - s_\mathrm{B} $, $ s = s_\mathrm{A} + s_\mathrm{B} $, $ A = J/a $, and $ K_u = 2K/a^3 $.

\begin{figure}[htbp]
	\includegraphics[width=1\columnwidth]{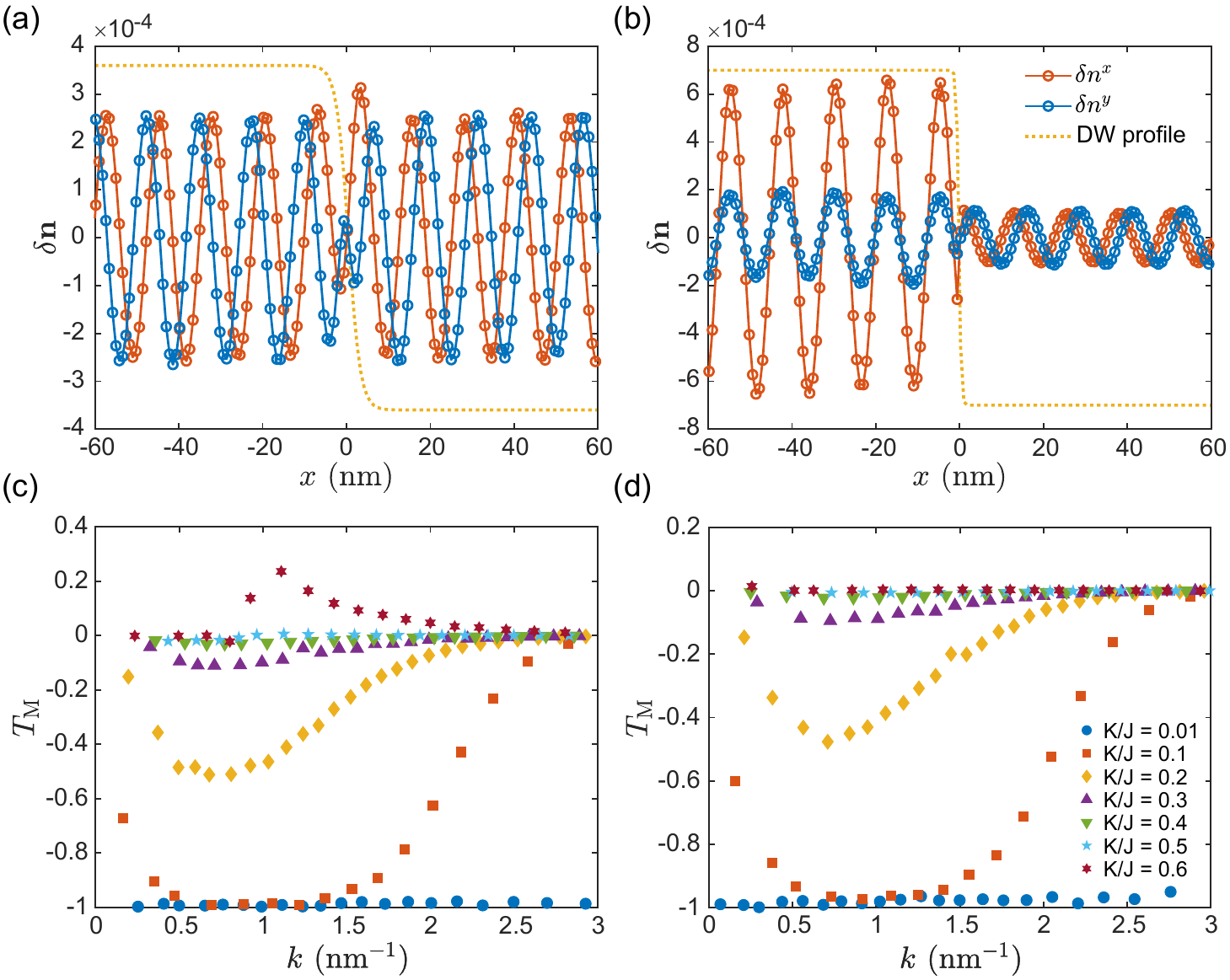}%
	\caption{\label{Fig:2} The distributions of the variation of the $ x $- and $ y $-component of N\'{e}el vector $\mathbf{n} $ under $ K/J = 0.01 $ (a) and $K/J =0.3 $ (b) for an incoming RH magnon with $ k = 0.5$ $\rm{nm^{-1}}$. $T_{\mathrm{M}}$ as a function of $k$ under different $K/J$ for an incoming RH (c) and LH (d) magnon. }
\end{figure}

In the local frame, we assume $ {\bf{n}} = (\delta n^X, \delta n^Y, 1) $ in Eq.~\eqref{eq:7} with $|\delta n^X|, |\delta n^Y| \ll 1$, which is justified for small-angle precession of local magnetic moments. Substituting the Walker ansatz $ \theta(x) = 2 \arctan [\mathrm{exp} (x/\Delta)]$ into Eq. \eqref{eq:7}, we obtain
\begin{equation}
	q^2 \varphi (\xi) = \Big (-\frac{d^2}{d \xi^2} - 2\mathrm{sech}^2 \xi\Big ) \varphi (\xi), \label{eq:8}
\end{equation}
where $\Delta = \sqrt{A/K_u} $ is the DW width, $ q^2 = (\sigma_s \omega^2 - \delta_s \omega) / {K_u}-1 $, $ \varphi = \delta n^X \pm i\delta n^Y $, and $ \xi = (x-x_0) / {\Delta} $. The hyperbolic potential $ V = - 2\mathrm{sech}^2 \xi $ \cite{yan2011all,kim2014propulsion} allows a total transmission of all magnon modes, as shown in Fig.~\ref{Fig:2}(a). We also observe that $ \delta n_x $ is ahead (behind) of $ \delta n_y $ in the left (right) domain, which clearly indicates an angular momentum switching of magnon, as schematically plotted in the middle panel of Fig.~(\ref{Fig:1}). Figure \ref{Fig:2}(b) shows that the amplitude of spin wave decreases when it passes through the domain wall for $ K/J = 0.3$, indicating that the practical scalar potential is no longer reflectionless \cite{SM}. Meanwhile, when an incident spin wave with the same wave vector passes through DW of different widths, there are different transmittances. As the width of DW decreases, the transmittance of the magnon decreases.

\begin{figure}[htbp]
	\includegraphics[width=1\columnwidth]{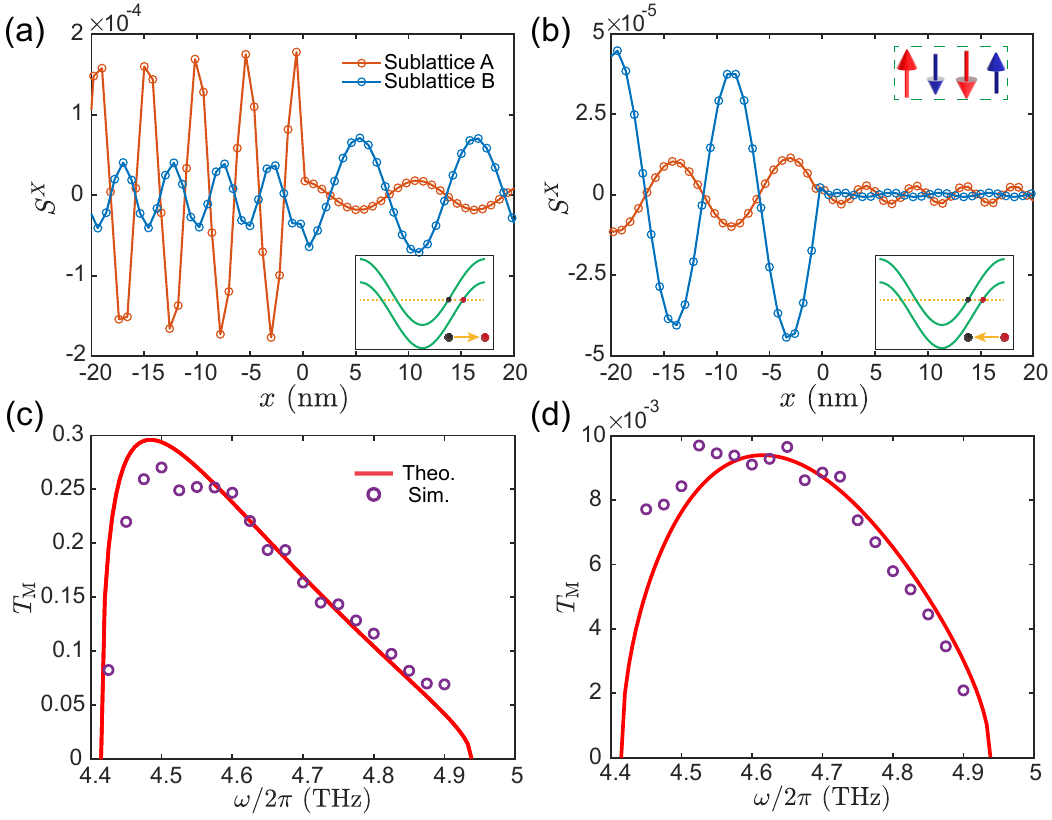}%
	\caption{\label{Fig:3} Spatial profiles of $ S^X $ with an incoming (a) RH and (b) LH spin wave at $\omega / 2\pi = 4.45$ THz with $ K/J = 1 $. Inset: green curves shows the magnon dispersion \cite{SM} where the black (red) dot indicates the spin wave in the left (right) side of sharp DW. The orange arrow shows the interband scattering direction. The dependence of $ T_{\mathrm{M}} $ on $ \omega/2\pi $ with an incoming (c) RH and (d) LH spin wave. Curves and circles represent the analytical results \eqref{eq:13} and numerical simulations, respectively.}
\end{figure}  

To quantitatively describe the scattering between spin waves and DWs, we define the transmittance based on $z$-component of magnon spin current $T_{\mathrm{M}} = J_z^{t}/J_{z}$, where $ J_z^{t} $ is the $z$-component of transmitted magnon spin current. In atomic lattices, the $z$-component of magnon spin current $ J_{z} = A \left( {\bf{n}} \times \nabla {\bf{n}} \right)_z $ reads
\begin{equation}
J_{z}= \pm\frac{J}{2a^2}[(\mathcal{R}_{s_\mathrm{A}})^2+(\mathcal{R}_{s_\mathrm{B}})^2-2\mathcal{R}_{s_\mathrm{A}}\mathcal{R}_{s_\mathrm{B}}\cos(ka)] \sin(2ka),\label{eq:12}
\end{equation}
where $\mathcal{R}_{s_\mathrm{A}}$ and $\mathcal{R}_{s_\mathrm{B}}$ are the amplitudes of magnon in each sublattice, and the RH (LH) magnon takes on the negative (positive) sign. Using the micromagnetic simulations, we determine the transmittance $ T_{\mathrm{M}} $ under different $ K/J $, as shown in Figs.~\ref{Fig:2}(c) and \ref{Fig:2}(d). The negative values of $ T_{\mathrm{M}} $ represent spin-wave chiral flipping after passing through the DW. Meanwhile, for $ K/J \leq 0.5 $, $T_{\mathrm{M}}$ is similar for an incoming RH and LH spin waves, indicating the insensitivity of polarization. However, in the case of $ K/J = 0.6 $, there exists noticeable difference between $T_{\mathrm{M}}$ of the RH and LH magnon, which results from the polarization-dependent interband magnon scattering and preserves the chirality of magnon (see analysis below).

\textit{The sharp DW.---}It is noted that, when $ K/J \geq 2/3 $, an abruptly sharp DW ($ \theta_{n} = 0$ or $\pi $) forms with the spins pointing towards the $ \pm z $ axis [see Eq.~\eqref{eq:15} below], which is plotted in the bottom panel of Fig.~(\ref{Fig:1}). Figures \ref{Fig:3}(a) and \ref{Fig:3}(b) show the profiles of $ S^X $ for an incoming RH and LH spin wave, respectively, with $\omega / 2\pi = 4.45$ THz under $K/J = 1$. Strikingly different from the case of wide DWs, we observe that the transmission coefficient sensitively depends on the polarization of incoming spin waves and the wave vector $ k $ varies when it passes through the DW. As shown in the inset of Fig.~\ref{Fig:3}(a) and \ref{Fig:3}(b), the interband magnon scattering induces the variation of magnon wave vector. Specifically, it is found that the amplitude of spin wave in sublattice A decreases, while it increases in sublattice B for an incoming RH spin wave. For the case of an LH spin wave, the amplitudes in both sublattices decrease after the transmission. To reach a quantitative understanding, we derive both the magnon reflection and transmission coefficient below.

First of all, the four spins in the dashed green box of the bottom panel of Fig.~(\ref{Fig:1}) [or the top-right inset in Fig.~(\ref{Fig:3})(b)] are used as the boundary condition. For an incoming RH spin wave, we obtain
\begin{equation}
	\begin{array}{c}
	(2K+s_\mathrm{B}\omega_\mathrm{+})\psi_{0}^{-} + J\psi_{-1}^{+} + J\psi_{1}^{-} = 0\\
	(2K+s_\mathrm{A}\omega_\mathrm{+})\psi_{1}^{-} + J\psi_{0}^{-} + J\psi_{2}^{+} = 0
	\end{array}. \label{eq:9}
\end{equation}
From the $ \psi_{0}^{-} $ and $ \psi_{1}^{-} $ in Eq.~\eqref{eq:9}, it is evident that the two spins at the domain interface automatically exhibit identical LH precession \cite{xing2022comparison}. Hence, the angular momentum of magnon remains when it passes through the sharp DW. However, due to the different dispersions for the same polarized spin wave in two sides of DW, there is an interband magnon scattering, associating with a linear momentum transfer. Meanwhile, it is worth noting that, because of the huge atomic pinning, the sharp DW cannot be driven by the linear force \cite{yang2019atomic}. For $ n \leq 0 $, $ \psi_{n}^{+} = \Xi_{s_\mathrm{A}}^{+} (e^{ikna} + r_\mathrm{R} e^{-ikna}) $ and $ \psi_{n}^{-} = \Xi_{s_\mathrm{B}}^{-} (e^{ik na} + r_\mathrm{R} e^{-ik na}) $. As for $ n>0 $, $ \psi_{n}^{-} = t_\mathrm{R} \Xi_{s_\mathrm{A}}^{-} e^{ik^{\prime}na} $ and $ \psi_{n}^{+} = t_\mathrm{R} \Xi_{s_\mathrm{B}}^{+} e^{ik^{\prime}na}$, where $ k $ and $ k^{\prime} $ are the wave vector of RH spin wave at the left and right sides of the sharp DW, respectively. We derive the reflection and transmission coefficients for RH magnons
\begin{equation}
r_\mathrm{R} = \frac{\mathcal{A} \mathcal{B}^{\ast}-J^2\rho^\mathrm{-}}{J^2\rho^\mathrm{-}-\mathcal{A} \mathcal{B}}, \label{eq:10}
\end{equation}
and
\begin{equation}
	t_\mathrm{R} = \frac{J^2\rho^\mathrm{+}(e^{ika}-e^{-ika})}{\beta(J^2\rho^\mathrm{-}-\mathcal{A} \mathcal{B})e^{ik^{\prime}a}}, \label{eq:11}
\end{equation}
where $  \beta = \Xi_{s_\mathrm{B}}^{-}/\Xi_{s_\mathrm{B}}^{+} $, $  \mathcal{A} = 2K\rho^\mathrm{-}+s_{\mathrm{A}} \omega_\mathrm{+} \rho^\mathrm{-}+Je^{ik^{\prime}a} $ and $ \mathcal{B} = 2K+s_{\mathrm{B}} \omega_\mathrm{+}+J\rho^\mathrm{+}e^{ik^{\prime}a} $. For a LH spin wave, one can do the following substitution: $ \omega_\mathrm{+} \rightarrow -\omega_\mathrm{-} $, $ \rho^\mathrm{+} \leftrightarrow \rho^\mathrm{-} $ in Eq.~\eqref{eq:11}.

In the case of an incident RH spin wave, the transmission of the $ z $-component of magnon spin current $ T_{\mathrm{M}} $ can be expressed as
\begin{equation}
	T_{\mathrm{M}} =  \frac{\left|\beta t_\mathrm{R}\right|^2[1+(\rho^-)^2-2\rho^-\cos(k^{\prime}a)]\sin(2k^{\prime}a)}{[1+(\rho^+)^2-2\rho^+\cos(ka)]\sin(2ka)} .\label{eq:13}
\end{equation}

\begin{figure}[htbp]
	\includegraphics[width=1\columnwidth]{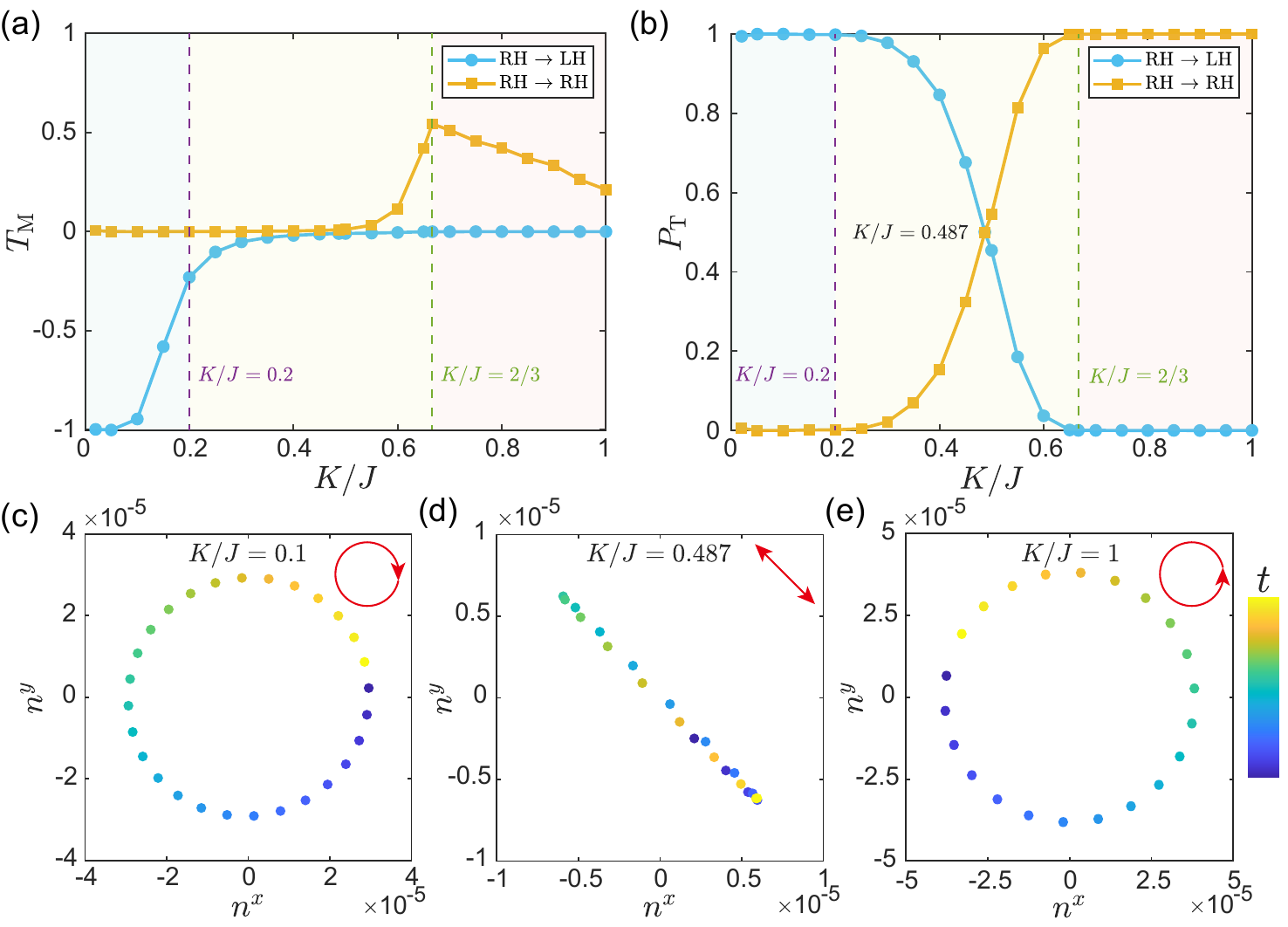}%
	\caption{\label{Fig:4} Dependence of $ T_{\mathrm{M}} $ (a) and $ P_{\mathrm{T}} $ (b) on $ K/J $ for an incident RH magnon with $ k = 2.0$ $\rm{nm^{-1}}$. Blue and orange curves represent the spin-wave chiral flipping and reserving process, respectively. The time evolution of the $ x $- and $ y $-components of N\'{e}el vector $\mathbf{n} $ under $ K/J = 0.1$ (c), $0.487$ (d), and $1.0$ (e) at $ x = 200 $ nm, respectively. The color bar shows the accumulation of the time.}
\end{figure}

It can be seen that the $ T_{\mathrm{M}} $ first increases and then decreases with the frequency, as illustrated in Figs.~\ref{Fig:3}(c) and \ref{Fig:3}(d). The two zero points result from the zero group velocity of the spin wave. Meanwhile, the RH magnon passes more easily than the LH one, which can be attributed to the distinct bound states of spin waves associated with the DW \cite{faridi2022atomic}. In addition, when the angular momentum densities $ s_\mathrm{A} $ and $ s_\mathrm{B} $ switch, there exist only slight differences of $ T_{\mathrm{M}} $ \cite{SM}.

To further understand the scattering between the magnon and DW, we fix the wavelength of the incident RH spin wave and adjust the DW width, as shown in Fig.~\ref{Fig:4}. It can be seen that the $ T_{\mathrm{M}} $ for spin wave chirality flipping process is negative and saturates to zero as the DW width decreases. However, for the chirality conserving process, $ T_{\mathrm{M}} $ reaches the maximum at $ K/J = 2/3 $ and then decreases. To interpret it, we consider fluctuations of a single domain \cite{barbara1994magnetization} as
\begin{equation}
	\mathcal{W} (\theta_{2n} - \pi) = \theta_{2n-1}+\theta_{2n+1}, \label{eq:14}
\end{equation}
where $ \mathcal{W} = 2(K/J+1) $. The following solution is found: $ \theta_{2n} = \theta_\mathrm{s} e^{-2n\phi}+\pi $ and $ \theta_{2n-1} = \theta_\mathrm{s} e^{-(2n-1)\phi}$, where $ \phi = \cosh^{-1}(\mathcal{W}/2) $ and $ \theta_\mathrm{s} $ is the tilted angle for the upward spin. Equation \eqref{eq:3} in sharp DW then can be recast as
\begin{equation}
	\frac{4}{3}\Big [\frac{K}{J}-1-\frac{(\lambda-1)^3}{8}\Big ]\theta_\mathrm{s}^2 = 2\frac{K}{J}-1-\lambda, \label{eq:15}
\end{equation}
where $ \lambda = e^{-\phi} $. The phase transition emerges from $ \theta_\mathrm{s} \neq 0 $ to $ \theta_\mathrm{s} = 0 $ with $ K/J = 2/3 $. To describe the ratio of two transmitted spin waves with different chiralities, we define the proportion $ P_{\mathrm{T}} = |T_{\mathrm{M}}^{s}|/\sum_s|T_{\mathrm{M}}^{s}| $, where $ T_{\mathrm{M}}^{s} $ represents the transmittance of the spin-wave chirality flipping or conserving process, as shown in Fig.~\ref{Fig:4}(b). For $ K/J \leq 0.2 $, the chirality flipping process is dominant, which transforms the incoming RH magnon to a LH one, as depicted in Fig.~\ref{Fig:4}(c). For $ K/J \geq 2/3 $, the spin-wave chirality conserving process is dominant, which reserves the chirality of magnon, as presented in Fig.~\ref{Fig:4}(e). In the region between the wide and sharp DW, the magnon follows a chiral asymptote. Noteworthily, at $ K/J = 0.487 $, we observe $ P_{\mathrm{T}}=1/2 $ that represents a zero magnon spin current, and corresponds to a linearly polarized magnon state, as shown in Fig.~\ref{Fig:4}(d). Hence, we can conveniently manipulate the chirality of magnon by tuning the DW width.

\begin{figure}[htbp]
	\includegraphics[width=1\columnwidth]{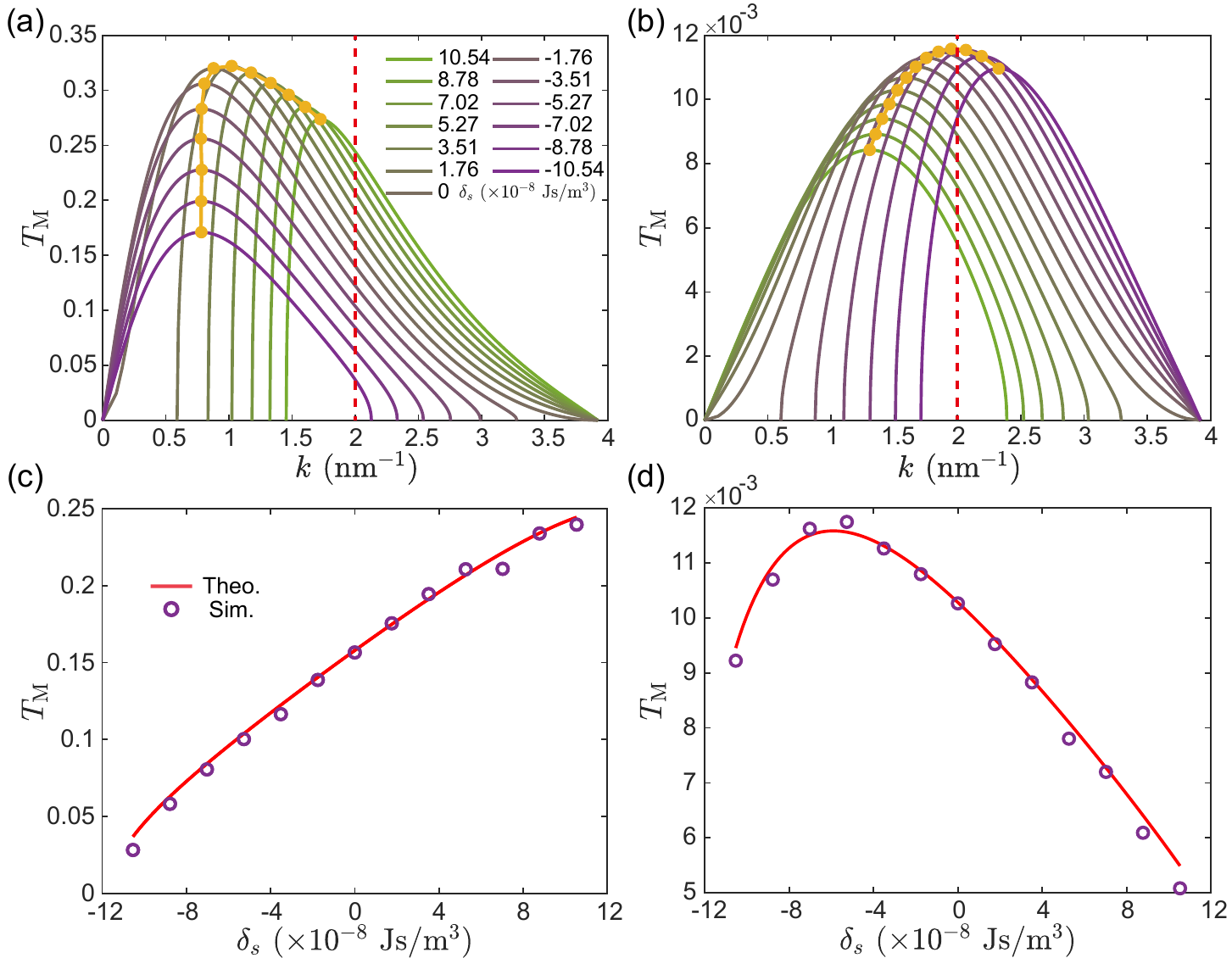}%
	\caption{\label{Fig:5} $ T_{\mathrm{M}} $ as a function of the wave vector $ k $ under different net angular momentum densities for an incident RH (a) and LH (b) magnons with $ K/J = 1 $. The orange dots label the maximum of each curve. $ T_{\mathrm{M}} $ at various net angular momentum densities $ \delta_s $ for an incident RH (c) and LH (d) magnons with $ k = 2.0 $ nm$ ^{-1} $. Red curves and purple circles show analytical results \eqref{eq:13} and numerical simulations, respectively.}
\end{figure}

In previous calculations, we have fixed the net spin angular momentum density $ \delta_s $ of FiMs. To further modulate the magnon transmittance, we vary $ \delta_s $ under a constant $ K/J = 1 $, as shown in Figs.~\ref{Fig:5}(a) and \ref{Fig:5}(b). As $ \delta_s $ increases, the maximum value of transmittance for an incoming RH and LH magnon first increases and then decreases, as indicated by the orange dots. Then, we fix the wave vector $ k = 2.0 $ nm$ ^{-1} $ of incident spin waves, which is indicated by the red dotted line in Figs.~\ref{Fig:5}(a) and \ref{Fig:5}(b). It can be seen that for an incoming RH magnon the transmittance increases as $\delta_s $ increases, see Fig.~\ref{Fig:5}(c). As for the incident LH magnon, the transmittance $ T_{\mathrm{M}} $ first increases and then decreases as $\delta_s $ increases, see Fig.~\ref{Fig:5}(d). Simulation results agree well with our analytical formula.

It is crucial to point out that the fundamental reason for the interband magnon scattering is merely $ \theta_{s} = 0$. Consequently, an artificially sharp DW, stabilized by the RKKY coupling, is sufficient induce a significant interband magnon scattering \cite{SM}. The roles of dipolar interaction and magnetoelastic coupling are interesting issues for future study. Generalizing the present formalism to higher dimensions is an open question.

In conclusion, we investigated the spin transport of magnons through atomic ferrimagnetic DWs. In the case of a wide DW, it was found that the magnon flips its spin, echoing its ferromagnetic and antiferromagnetic counterparts. However, for sharp DWs, the magnon reserves the spin due to the emerging interband magnon scattering. In the intermediate region, the transmission magnons with two spins mix and can even carry zero magnon spin current under some circumstances. Moreover, we identified that the net angular momentum density of FiMs can effectively modulate the magnon transmittance. These findings may greatly advance our understanding of the scattering between magnons and magnetic textures in FiMs.

\begin{acknowledgments}
We thank Z. Jin and Y. Cao for useful discussions. This work was funded by the National Key R$\&$D Program under Contract No. 2022YFA1402802 and the National Natural Science Foundation of China (NSFC) (Grants No. 12374103 and No. 12074057).
\end{acknowledgments}

\begin{thebibliography}{99}
	\newcommand{\DOI}[1]{doi: \href{https://doi.org/#1}{#1}}
	
	\bibitem{serga2010yig}A. Serga, A. Chumak, and B. Hillebrands, YIG magnonics, \href{https://doi.org/10.1088/0022-3727/43/26/264002}{J. Phys. D \textbf{43}, 264002 (2010).}
	
	\bibitem{chumak2015magnon}A. V. Chumak, V. I. Vasyuchka, A. A. Serga, and
	B. Hillebrands, Magnon spintronics, \href{https://doi.org/10.1038/nphys3347}{Nat. Phys. \textbf{11}, 453 (2015).}
	
	\bibitem{yu2021magnetic}H. Yu, J. Xiao, and H. Schultheiss, Magnetic texture based magnonics, \href{https://doi.org/10.1016/j.physrep.2020.12.004}{Phys. Rep. \textbf{905}, 1 (2021).}
	
	\bibitem{barman20212021}A. Barman, G. Gubbiotti, S. Ladak, A. O. Adeyeye, M. Krawczyk, J. Gr{\"a}fe, C. Adelmann, S. Cotofana, A. Naeemi, V. I. Vasyuchka, \emph{et al}., The 2021 magnonics roadmap, \href{https://doi.org/10.1088/1361-648X/abec1a}{J. Phys.: Condens. Matter \textbf{33}, 413001 (2021).}
	
	\bibitem{yuan2022quantum}H. Yuan, Y. Cao, A. Kamra, R. A. Duine, and P. Yan, Quantum magnonics: When magnon spintronics meets quantum information science, \href{https://doi.org/10.1016/j.physrep.2022.03.002}{Phys. Rep. \textbf{965}, 1 (2022).}
	
	\bibitem{yan2011all}P. Yan, X. S. Wang, and X. R. Wang, All-magnonic spin-transfer torque and domain wall propagation, \href{https://doi.org/10.1103/PhysRevLett.107.177207}{Phys. Rev. Lett. \textbf{107}, 177207 (2011).}
	
	\bibitem{wang2012domain}X. S. Wang, P. Yan, Y. Shen, G. E. Bauer, and X. R. Wang, Domain wall propagation through spin wave emission, \href{https://doi.org/10.1103/PhysRevLett.109.167209}{Phys. Rev. Lett. \textbf{109}, 167209 (2012).}
	
	\bibitem{yu2016magnetic}W. Yu, J. Lan, R. Wu, and J. Xiao, Magnetic Snell's law and spin-wave fiber with Dzyaloshinskii-Moriya interaction, \href{https://doi.org/10.1103/PhysRevB.94.140410}{Phys. Rev. B \textbf{94}, 140410(R) (2016).}
	
	\bibitem{hamalainen2018control}S. J. H{\"a}m{\"a}l{\"a}inen, M. Madami, H. Qin, G. Gubbiotti, and
	S. van Dijken, Control of spin-wave transmission by a programmable domain wall, \href{https://doi.org/10.1038/s41467-018-07372-x}{Nat. Commun. \textbf{9}, 4853 (2018).}
	
	\bibitem{oh2019bidirectional}S.-H. Oh, S. K. Kim, J. Xiao, and K.-J. Lee, Bidirectional spin-wave-driven domain wall motion in ferrimagnets, \href{https://doi.org/10.1103/PhysRevB.100.174403}{Phys. Rev. B \textbf{100}, 174403 (2019).}
	
	\bibitem{han2019mutual}J. Han, P. Zhang, J. T. Hou, S. A. Siddiqui, and L. Liu, Mutual control of coherent spin waves and magnetic domain walls in a magnonic device, \href{https://doi.org/10.1126/science.aau2610}{Science \textbf{366}, 1121 (2019).}
	
	\bibitem{liang2022nonreciprocal}X. Liang, Z. Wang, P. Yan, and Y. Zhou, Nonreciprocal spin waves in ferrimagnetic domain-wall channels, \href{https://doi.org/10.1103/PhysRevB.106.224413}{Phys. Rev. B \textbf{106}, 224413 (2022).}
	
	\bibitem{lan2017antiferromagnetic}J. Lan, W. Yu, and J. Xiao, Antiferromagnetic domain wall as spin wave polarizer and retarder, \href{https://doi.org/10.1038/s41467-017-00265-5}{Nat. Commun. \textbf{8}, 178 (2017).}
	
	\bibitem{faridi2022atomic}E. Faridi, S. K. Kim, and G. Vignale, Atomic-scale spin-wave polarizer based on a sharp antiferromagnetic domain wall, \href{https://doi.org/10.1103/PhysRevB.106.094411}{Phys. Rev. B \textbf{106}, 094411 (2022).}
	
	\bibitem{iwasaki2014theory}J. Iwasaki, A. J. Beekman, and N. Nagaosa, Theory of magnon-skyrmion scattering in chiral magnets, \href{https://doi.org/10.1103/PhysRevB.89.064412}{Phys. Rev. B \textbf{89}, 064412 (2014).}
	
	\bibitem{schutte2014magnon}C. Sch{\"u}tte and M. Garst, Magnon-skyrmion scattering in chiral magnets, \href{https://doi.org/10.1103/PhysRevB.90.094423}{Phys. Rev. B \textbf{90}, 094423 (2014).}
	
	\bibitem{kim2019tunable}S. K. Kim, K. Nakata, D. Loss, and Y. Tserkovnyak, Tunable magnonic thermal hall effect in skyrmion crystal phases of ferrimagnets, \href{https://doi.org/10.1103/PhysRevLett.122.057204}{Phys. Rev. Lett. \textbf{122}, 057204 (2019).}
	
	\bibitem{wang2021magnonic}Z. Wang, H. Yuan, Y. Cao, Z.-X. Li, R. A. Duine, and P. Yan, Magnonic frequency comb through nonlinear magnon-skyrmion scattering, \href{https://doi.org/10.1103/PhysRevLett.127.037202}{Phys. Rev. Lett. \textbf{127}, 037202 (2021).}
	
	\bibitem{li2022interaction}Z. Li, M. Ma, Z. Chen, K. Xie, and F. Ma, Interaction between magnon and skyrmion: Toward quantum magnonics, \href{https://doi.org/10.1063/5.0121314}{J. Appl. Phys. \textbf{132}, 210702 (2022).}
	
	\bibitem{schryer1974motion}N. L. Schryer and L. R. Walker, The motion of 180{$^\circ$} domain walls in uniform dc magnetic fields, \href{https://doi.org/10.1063/1.1663252}{J. Appl. Phys. \textbf{45}, 5406 (1974).}
	
	\bibitem{wang2018theory} X. S. Wang, H. Y. Yuan, and X. R. Wang, A theory on skyrmion size, \href{https://doi.org/10.1038/s42005-018-0029-0}{Commun. Phys. \textbf{1}, 31 (2018).}
	
	\bibitem{novoselov2003subatomic}K. S. Novoselov, A. K. Geim, S. V. Dubonos, E. W. Hill, and I. V. Grigorieva, Subatomic movements of a domain wall in the Peierls potential, \href{https://doi.org/10.1038/nature02180}{Nature \textbf{426}, 812 (2003).}
	
	\bibitem{yan2012magnonic}P. Yan and G. E. Bauer, Magnonic domain wall heat conductance in ferromagnetic wires, \href{https://doi.org/10.1103/PhysRevLett.109.087202}{Phys. Rev. Lett. \textbf{109}, 087202 (2012).}
	
	\bibitem{Yang2019}H. Yang, H. Y. Yuan, M. Yan, H. W. Zhang, and P. Yan, Atomic antiferromagnetic domain wall propagation beyond the relativistic limit, \href{https://doi.org/10.1103/PhysRevB.100.024407}{Phys. Rev. B \textbf{100}, 024407 (2019).}
	
	\bibitem{krizek2022atomically}F. Krizek, S. Reimers, Z. Ka{\v{s}}par, A. Marmodoro, J. Michali{\v{c}}ka, O. Man, A. Edstr{\"o}m, O. J. Amin, K. W. Edmonds, R. P. Campion, \emph{et al}., Atomically sharp domain walls in an antiferromagnet, \href{https://doi.org/10.1126/sciadv.abn3535}{Sci. Adv. \textbf{8}, eabn3535 (2022).}
	
	\bibitem{lee2023giant}Y. Lee, S. Son, C. Kim, S. Kang, J. Shen, M. Kenzelmann,
	B. Delley, T. Savchenko, S. Parchenko, W. Na, \emph{et al}., Giant magnetic anisotropy in the atomically thin van der Waals antiferromagnet FePS$_3$, \href{https://doi.org/10.1038/s41563-020-0713-9}{Adv. Electron. Mater. \textbf{9}, 2200650 (2023).}
	
	\bibitem{tveten2014antiferromagnetic}E. G. Tveten, A. Qaiumzadeh, and A. Brataas, Antiferromagnetic domain wall motion induced by spin waves, \href{https://doi.org/10.1103/PhysRevLett.112.147204}{Phys. Rev. Lett. \textbf{112}, 147204 (2014).}
	
	\bibitem{kim2014propulsion}S. K. Kim, Y. Tserkovnyak, and O. Tchernyshyov, Propulsion of a domain wall in an antiferromagnet by magnons, \href{https://doi.org/10.1103/PhysRevB.90.104406}{Phys. Rev. B \textbf{90}, 104406 (2014).}
	
	\bibitem{kim2022ferrimagnetic}S. K. Kim, G. S. Beach, K.-J. Lee, T. Ono, T. Rasing,
	and H. Yang, Ferrimagnetic spintronics, \href{https://doi.org/10.1038/s41563-021-01139-4}{Nat. Mater. \textbf{21}, 24 (2022).}
	
	\bibitem{zhang2023ferrimagnets}Y. Zhang, X. Feng, Z. Zheng, Z. Zhang, K. Lin, X. Sun,
	G. Wang, J. Wang, J. Wei, P. Vallobra, \emph{et al}., Ferrimagnets for spintronic devices: From materials to applications, \href{https://doi.org/10.1063/5.0104618}{Appl. Phys. Rev. \textbf{10}, 011301 (2023).}
	
	\bibitem{oh2017coherent}S.-H. Oh, S. K. Kim, D.-K. Lee, G. Go, K.-J. Kim,
	T. Ono, Y. Tserkovnyak, and K.-J. Lee, Coherent terahertz spin-wave emission associated with ferrimagnetic domain wall dynamics, \href{https://doi.org/10.1103/PhysRevB.96.100407}{Phys. Rev. B \textbf{96}, 100407(R) (2017).}
	
	\bibitem{kim2021current}D.-H. Kim, S.-H. Oh, D.-K. Lee, S. K. Kim, and K.-J.
	Lee, Current-induced spin-wave Doppler shift and attenuation in compensated ferrimagnets, \href{https://doi.org/10.1103/PhysRevB.103.014433}{Phys. Rev. B \textbf{103}, 014433 (2021).}
	
	\bibitem{kim2020distinct}C. Kim, S. Lee, H.-G. Kim, J.-H. Park, K.-W. Moon, J. Y. Park, J. M. Yuk, K.-J. Lee, B.-G. Park, S. K. Kim, \emph{et al}., Distinct handedness of spin wave across the compensation temperatures of ferrimagnets, \href{https://doi.org/10.1038/s41563-020-0722-8}{Nat. Mater. \textbf{19}, 980 (2020).}
	
	\bibitem{okamoto2020flipping}S. Okamoto, Flipping handedness in ferrimagnets, \href{https://doi.org/10.1038/s41563-020-0779-4}{Nat. Mater. \textbf{19}, 929 (2020).}
	
	\bibitem{li2023puzzling}R. Li, L. J. Riddiford, Y. Chai, M. Dai, H. Zhong, B. Li,
	P. Li, D. Yi, Y. Zhang, D. A. Broadway, \emph{et al}., A puzzling insensitivity of magnon spin diffusion to the presence of 180-degree domain walls, \href{https://doi.org/10.1038/s41467-023-38095-3}{Nat. Commun. \textbf{14}, 2393 (2023).}
	
	\bibitem{SM}See Supplemental Material at http://link.aps.org/supplemental/ for materials parameters adopted in calculations, the derivation of the dispersion of spin wave in ferrimagnets, the magnon transport through sharp FiM domain wall by switching $ s_\mathrm{A} $ and $ s_\mathrm{B} $, and the influence of interfacial exchange interaction $ J_{12} $.
	
	
	\bibitem{xing2022comparison}Y. W. Xing, Z. R. Yan, and X. F. Han, Comparison of spin-wave transmission in parallel and antiparallel magnetic configurations, \href{https://doi.org/10.1103/PhysRevB.105.064427}{Phys. Rev. B \textbf{105}, 064427 (2022).}
	
	\bibitem{yang2019atomic}H. Yang, H. Yuan, M. Yan, H. Zhang, and P. Yan, Atomic antiferromagnetic domain wall propagation beyond the relativistic limit, \href{https://doi.org/10.1103/PhysRevB.100.024407}{Phys. Rev. B \textbf{100}, 024407 (2019).}
	
	\bibitem{barbara1994magnetization}B. Barbara, Magnetization processes in high anisotropy systems, \href{https://doi.org/10.1016/0304-8853(94)90432-4}{J. Magn. Magn. Mater. \textbf{129}, 79 (1994).}
	
	

\end{thebibliography}

\end{document}